\documentclass{article}

\usepackage{amssymb}
\usepackage[all]{xy} 
 \usepackage{rotating}
 \usepackage{fancybox}

 \newcommand{\variable}{\bullet}

\usepackage{theorem}
\newtheorem{thm}{Theorem}[section]
\newtheorem{prop}[thm]{Proposition}
\newtheorem{coro}[thm]{Corollary}

\theorembodyfont{\normalfont}
\newtheorem{defn}[thm]{Definition}
\newtheorem{exam}[thm]{Example}

\newenvironment{dproof}{\noindent\textit{Proof. }}{$\square$\\ \null}
\newenvironment{proofof}[1]{\noindent\textit{Proof of #1. }}{$\square$\\ \null}

\newcommand{\emptystr}{\epsilon}
\newcommand{\sw}{SW}    
\newcommand{\lb}{\!:\!}
\newcommand{\Om}{\Omega}
\newcommand{\edge}[1]{[#1]} 
\newcommand{\suc}{\mathit{succ}}
\newcommand{\ari}{\mathrm{ar}}
\newcommand{\bN}{\mathbb{N}} 
\newcommand{\cN}{\mathcal{N}} 
\newcommand{\cL}{\mathcal{L}} 
\newcommand{\cS}{\mathcal{S}} 
\newcommand{\cX}{\mathcal{X}} 
\newcommand{\Gr}{\mathbf{Gr}} 
\newcommand{\Set}{\mathbf{Set}}

\title{Data-Structure Rewriting}

\author{D. Duval$^1$, R. Echahed$^2$, F. Prost$^2$ \\
 Institut d'Informatique et de Math\'ematiques 
            Appliqu\'ees de Grenoble\\
        \{Dominique.Duval $|$ Rachid.Echahed $|$ Frederic.Prost\}@imag.fr \\
  $\ ^1$ Laboratoire LMC, B. P. 53, 38041 Grenoble, France \\
$\ ^2$ Laboratoire LEIBNIZ,
46, av. Felix Viallet,
38031  Grenoble, France}

\begin{document}

\maketitle

\vspace{-0.5cm}
\begin{abstract}
We tackle the problem of data-structure rewriting including pointer
redirections. We propose two basic rewrite steps: (i) Local
Redirection and Replacement steps the aim of which is redirecting
specific pointers determined by means of a pattern, as well as adding
new information to an existing data ; and (ii) Global Redirection
steps which are aimed to redirect all pointers targeting a node
towards another one. We define these two rewriting steps following the
double pushout approach. We define first the category of graphs we
consider and then define rewrite rules as pairs of graph homomorphisms
of the form $L \leftarrow K \rightarrow R$.  Unfortunately, inverse
pushouts (complement pushouts) are not unique in our setting and
pushouts do not always exist. Therefore, we define rewriting steps so
that a rewrite rule can always be performed once a matching is found.
\end{abstract}
\section{Introduction}

Rewriting techniques have been proven to be very useful to establish
formal bases for high level programming laguages as well as theorem
provers. These techniques have been widely investigated for strings
\cite{BookF93}, trees or terms \cite{BaaderN98} and term graphs
\cite{Plu98a,BS99}.

In this paper we tackle the problem of rewriting classical
data-structures such as circular lists, double-chained lists, etc.
Even if such data-structures can be easily simulated by string or tree
processing, they remain very useful in designing algorithms with good
complexity. The investigation of data-structure rewrite systems will
contribute to define a clean semantics and proof techniques for
``pointer'' handling. It will also provide a basis for multiparadigm
programming languages integrating declarative (functional and logic)
and imperative features.

General frameworks of graph transformation are now well established,
see e.g. \cite{handbook1,handbook2,handbook3}. Unfortunately,
rewriting classical data-structures represented as cyclic graphs did
not benefit yet of the same effort as for terms or term graphs. Our
aim in this paper is to investigate basic rewrite steps for
data-structure transformation.  It turns out that pointer redirection
is the key issue we had to face, in addition to classical replacement
and garbage collection.  We distinguish two kinds of redirections:
(i)\emph{Global redirection} which consists in redirecting in a row
all edges pointing to a given node, to another node ; and (ii)
\emph{Local redirection} which consists in redirecting a particular
pointer, specified e.g. by a pattern, in order to point to a new
target node. Global redirection is very often used in the
implementation of functional programming languages, for instance when
changing roots of term graphs.  As for local redirection, it is useful
to express classical imperative algorithms.

We introduce two kind of rewrite steps. The first is one called
\emph{local redirection and replacement} and the second kind is dedicated
to global redirection. We define these steps following the double
pushout approach \cite{CorradiniMREHL97,HMD01}. We have chosen this
approach because it simplifies drastically the presentation of our
results. The algorithmic fashion, which we followed first, turns out
to be arduous. Thus, basic rewrite rules are given by a pair of graph
homomorphisms $L \leftarrow K \rightarrow R$. We precise the
r\^ole that plays $K$ in order to perform local or global redirection
of pointers.  The considered homomorphisms are not necessarily
injective in our setting, unlike classical assumptions as in the
recent proposals dedicated to graph programs
\cite{PlumpS04,HabelP01}. This means that inverse pushouts (complement
pushouts) are not unique.

The paper is organized as follows: The next section introduces the
category of graphs which we consider in the paper. Section~3 states
some technical results that help defining rewrite steps. Section~4
introduces data-structure rewriting and defines mainly two rewrite
steps, namely LRR-rewriting and GR-rewriting. We compare our proposal
to related work in section~5. Concluding remarks are given in
section~6.  Proofs are found in the appendix. We assume the reader is
familiar with basic notions of category theory (see e.g. \cite{AL} for an introduction).

\section{Graphs}

In this section we introduce the category of graphs we consider in the
paper.  These graphs are supposed to represent data-structures. We
define below such graphs in a mono-sorted setting. Lifting our results
to the  many-sorted case is straightforward.

\begin{defn}[Signature]
A \emph{signature} $\Om$ is a set of operation symbols such that
each operation symbol in $\Om$, say $f$, is provided by a natural number, $n$,
representing its \emph{arity}. We write $\ari(f) = n$. 
\end{defn}

In the sequel, we use the following notations.  Let $A$ be a
set. We note $A^*$ the set of strings made of elements in $A$.  Let $f
: A \to B$ be a function. We note $f^* : A^* \to B^*$ the unique
extension of $f$ over strings defined by $f^*(\emptystr) = \emptystr$
where $\emptystr$ is the empty string and $ f^*(a_1 \ldots a_n) =
f(a_1) \ldots f(a_n)$.

We assume that $\Om$ is fixed throughout the rest of the paper.

\begin{defn}[Graph]
A \emph{graph} $G$ is made of:
\begin{itemize}
\item a set of \emph{nodes} $\cN_G$, 
\item a subset of \emph{labeled nodes} $\cN_G^{\Om}\subseteq\cN_G$, 
\item a \emph{labeling function} $\cL_G:\cN_G^{\Om}\to\Om$,
\item and a \emph{successor function}
$\cS_G:\cN_G^{\Om}\to\cN_G^*$, 
\end{itemize}
such that,
for each labeled node $n$, the length of the string $\cS_G(n)$ 
is the arity of the operation $\cL_G(n)$.
\end{defn}

This definition can be illustrated by the following diagram,
where $\lg(u)$ is the length of the string $u$.
:
$$ \xymatrix{
\cN_G  & \cN_G^{\Om} \ar[l]_{\supseteq} \ar[d]_{\cL_G} \ar[r]^{\cS_G} 
  \ar@{}[dr]|{=} & \cN_G^*  \ar[d]^{\lg} \\ 
& \Om \ar[r]_{\ari} & \bN \\
}$$

Moreover: 
\begin{itemize}
\item the \emph{arity}
of a node $n$ is defined as the arity of its label,
\item the $i$-th successor of a node $n$ is denoted $\suc_G(n,i)$,
\item the \emph{edges} of a graph $G$ are the pairs 
$(n,i)$ where $n\in\cN_G^{\Om}$ and $i\in\{1,\dots,\ari(n)\}$,
the \emph{source} of an edge $(n,i)$ is the node $n$,
and its \emph{target} is the node $\suc_G(n,i)$,
\item the fact that $f=\cL_G(n)$ can be written as $n\lb f \,$,
\item the set of unlabeled nodes of $G$ is denoted
$\cN_G^{\cX}$, so that:
  $ \cN_G=\cN_G^{\Om}+\footnote{$+$ stands for disjoint union.}\cN_G^{\cX} \;.$
\end{itemize}

\begin{exam}
  \label{graphexample}
Let $G$ be the graph defined by 
 \begin{itemize}
  \item $\cN_G=\{m;n;o;p;q;r\}$
  \item $\cN_G^{\Om} = \{m;o;p\}$
  \item $\cN_G^{\cX} = \{n;q;r\}$
  \item $\cL_G$ is defined by: $[m \mapsto f; o \mapsto g; p \mapsto h]$ 
  \item $\cS_G$ is defined by:
          $[m \mapsto no; o \mapsto np; p \mapsto qrm]$
 \end{itemize}

  Graphically we represent this graph as:
 $
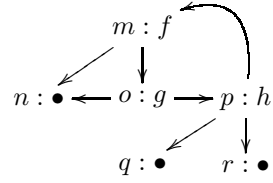
\xymatrix@R=1pc@C=1pc
     {&m:f\ar[dl] \ar[d]\\
      n:\variable& o:g \ar[l] \ar[r] & p:h \ar[d] \ar[dl] \ar@/^-2pc/[ul] \\
      & q:\variable & r:\variable}$

  We use $\variable$ to denote lack of label. Informally, one may
think of $\variable$ as anonymous variables.
\end{exam}

\begin{defn}[Graph homomorphism]
\label{defn:grhom}
A \emph{graph homomorphism} $\varphi:G\to H$
is a map $\varphi:\cN_G\to\cN_H$ such that
$\varphi(\cN_G^{\Om})$ is included in $\cN_H^{\Om}$
and, for each node $n\in\cN_G^{\Om}$:
  $ \cL_H(\varphi(n)) = \cL_G(n) \;\mbox{ and }\;
  \cS_H(\varphi(n)) = \varphi^*(\cS_G(n))\;.$
\end{defn}

Let $\varphi^{\Om}:\cN_G^{\Om}\to\cN_H^{\Om}$ denote the restriction
of $\varphi$ to the subset $\cN_G^{\Om}$. Then, the properties in the
definition above mean that the following diagrams are commutative: $$
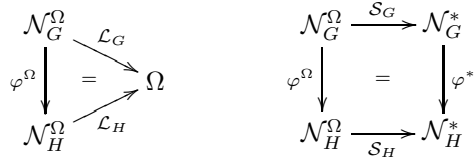
\xymatrix@R=.5pc{
\cN_G^{\Om} \ar[dd]_{\varphi^{\Om}} \ar[rd]^{\cL_G} & \\ 
\mbox{} \ar@{}[r]|(.4){=} & \Om \\ 
\cN_H^{\Om} \ar[ru]_{\cL_H} & \\ 
}
\qquad\qquad
\xymatrix{
\cN_G^{\Om} \ar[d]_{\varphi^{\Om}} \ar[r]^{\cS_G} \ar@{}[dr]|{=} 
  & \cN_G^* \ar[d]^{\varphi^*} \\ 
\cN_H^{\Om} \ar[r]_{\cS_H} & \cN_H^* \\ 
}
$$

The image $\varphi(n,i)$ of an edge $(n,i)$ 
of $G$ is defined as the edge $(\varphi(n),i)$ of $H$.
 
\begin{exam}
  \label{exam:hom}
  Consider the following graph $H$:
  \xymatrix@R=1pc@C=1pc{a:f \ar[d] \ar[r] & c:g \ar[r]\ar[d] &e:\variable\\
            b:\variable & d:\variable 
   }

 Let $\varphi : \cN_H\to\cN_G$, where $G$ is the graph defined in
Example~\ref{graphexample}, be defined as: $[a \mapsto m; b \mapsto n;
c\mapsto o; d \mapsto n; e \mapsto p]$. Map $\varphi$ is  a graph
homomorphism from $H$ to $G$. Notice that the nodes without
labels act as placeholders for any graph.
\end{exam}

It is easy to check that the graphs (as objects)
together with the graph homomorphisms (as arrows) form a
category, which is called the \emph{category of graphs} and noted $\Gr$ .

\section{Disconnected graphs and homomorphisms}
This section is dedicated to some technical definitions the aim of which is
the simplification of the definition of rewrite rules given in the following 
section.

\begin{defn}[Disconnected edge] 
An edge $(n,i)$ of a graph $G$ is \emph{disconnected} 
if its target $\suc_G(n,i)$ is unlabeled.
\end{defn}

The next definition introduces the notion of what we call disconnected
graph.  Roughly speaking, the disconnected graph associated to a
graph $G$ and a set of edges $E$ is obtained by redirecting every edge
in $E$ (whether it is yet disconnected or not) towards a \emph{new},
unlabeled, target.

\begin{defn}[Disconnected graph] 
The \emph{disconnected graph} associated to a graph $G$ 
and a set of edges $E$ of $G$
is the following graph $D(G,E)$:
\begin{itemize}
\item $\cN_{D(G,E)}=\cN_G+\cN_E$,
where $\cN_E$ is made of one new node $n\edge{i}$ for each 
edge $(n,i)\in E$,
\item $\cN_{D(G,E)}^{\Om}=\cN_G^{\Om}$,
\item for each $n\in \cN_G^{\Om}$: $\cL_{D(G,E)}(n)=\cL_G(n)$,
\item for each $n\in \cN_G^{\Om}$ and $i\in\{1,\dots,\ari(n)\}$: 
  \begin{itemize}
  \item if $(n,i)\not\in E$ then $\suc_{D(G,E)}(n,i)=\suc_G(n,i)$, 
  \item if $(n,i)\in E$ then $\suc_{D(G,E)}(n,i)=n\edge{i}$.
  \end{itemize}
\end{itemize}
\end{defn}

\begin{defn}[Connection homomorphism] 
The \emph{connection homomorphism} associated to a graph $G$ 
and a set of edges $E$ of $G$
is the homomorphism $\delta_{G,E}:D(G,E)\to G$ such that:
\begin{itemize}
\item if $n\in\cN_G$ then $\delta_{G,E}(n)=n$,
\item if $n\edge{i}\in\cN_E$ then $\delta_{G,E}(n\edge{i})=\suc_G(n,i)$.
\end{itemize}
\end{defn}

It is easy to check that $\delta_{G,E}$ is a graph homomorphism.

\begin{defn}[Disconnected homomorphism] 
The \emph{disconnected graph homomorphism} associated to a graph
homomorphism $\varphi:G\to H$
and a set of edges $E$ of $G$
is the homomorphism $D_{\varphi,E}:D(G,E)\to D(H,\varphi(E))$ 
defined as follows:
\begin{itemize}
\item if $n\in\cN_G$ then $D_{\varphi,E}(n)=\varphi(n)$,
\item if $n\edge{i}\in\cN_E$ then 
$D_{\varphi,E}(n\edge{i})=\varphi(n)\edge{i}$.
\end{itemize}
\end{defn}

It is easy to check that $D_{\varphi,E}$ is a graph homomorphism.

\begin{exam}
  \label{exam:deco_hom}

  Consider the graph $H$ of Example~\ref{exam:hom}. Then 
the disconnected graph associated to $H$ and the set of edges $\{(a,2);(c,1)\}$ 
is the following graph:

 $$ 
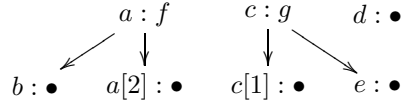
\xymatrix@R=1pc@C=1pc{& a:f \ar[dl] \ar[d]& c:g \ar[dr]\ar[d] & d:\variable \\
            b:\variable & a[2]:\variable& c[1]:\variable& e:\variable  
   }$$

  Note that even if edge $(c,1)$ is already disconnected in $H$ it is
\emph{redirected} towards a new unlabeled node,$c[1]$, in $D(H, \{(a,2) ;
(c,1)\})$.  

  Now if we consider the graph homomorphism $\varphi : H \to G $
defined in Example~\ref{exam:hom}, the disconnected graph homomorphism
$D_{\varphi,\{(a,2);(c,1)\}}: D(H,\{(a,2);(c,1)\}) \to
D(G,\{(m,2);(o,1)\})$ is the mapping 
 $[a \mapsto m; b \mapsto n;c\mapsto o; d \mapsto n; e \mapsto p;
a[2] \mapsto m[2]; c[1] \mapsto o[1]]$
\end{exam}

\section{Data-structure rewriting}

In this section we define data structure rewriting as a succession of 
rewrite steps.
A rewrite step is defined from a rewrite rule and a
matching.
A rewrite rule is a \emph{span} of graphs, i.e., 
a pair of graph homomorphisms with a common source: 
  $$ \xymatrix@C=5pc{ L & K \ar[l]_{\delta} \ar[r]^{\rho} & R }$$
A matching is a morphism of graphs:
  $ \xymatrix@R=2pc{ L \ar[r]^{\mu} & G }$.
There are two kinds of rewrite steps.
\begin{itemize}
\item The first kind is called \emph{Local Redirection and Replacement Rewriting}
(LRR-rewriting, for short). 
Its r\^ole is twofold: adding to $G$ a copy
of the instance of the right-hand side $R$, 
and performing some local redirections of edges specified by means
of the rewrite rule.
\item The second kind of rewrite steps is called \emph{Global Redirection Rewriting}
(GR-Rewriting, for short).
Its r\^ole consists in performing
redirections: all incoming edges of some node $a$ in $G$ 
are redirected to a node $b$.
\end{itemize}
We define LRR-rewriting and GR-rewriting in the two following subsections.
We use in both cases the double-pushout approach to define rewrite steps.

\subsection{LRR-rewriting}
 
Before defining LRR-rewrite rules and steps, we state first a technical
result about the existence of inverse pushouts in our setting.

\begin{thm}[An inverse pushout]
\label{thm:invpo}
Let $\mu:L\to U$ be a graph homomorphism, 
$E$ a set of edges of $L$,
and let $D_{\mu,E}:D(L,E)\to D(U,\mu(E))$
be the disconnected graph homomorphism associated to $\mu$
and $E$.
Then the following square is a pushout in the category of graphs ($\Gr$):
$$ \xymatrix@C=5pc{
L \ar[d]_{\mu} & D(L,E) \ar[d]^{D_{\mu,E}} \ar[l]_{\delta_{L,E}} \\ 
U & D(U,\mu(E)) \ar[l]^{\delta_{U,\mu(E)}} \\ 
}$$
\end{thm}

\begin{dproof} 
This result is an easy corollary of Theorem~\ref{thm:grsetpo}.
\end{dproof}

\begin{defn}[Disconnecting pushout]
Let $\mu:L\to U$ be a graph homomorphism 
and $E$ a set of edges of $L$.
The \emph{disconnecting pushout associated to $\mu$ and $E$} 
is the pushout from Theorem~\ref{thm:invpo}.
\end{defn}
 
It can be noted that the disconnecting pushout is not unique,
in the sense that there are generally several inverse 
pushouts of:
$$ \xymatrix@C=5pc{
L \ar[d]_{\mu} & D(L,E) \ar[l]_{\delta_{L,E}} \\ 
U &  \\ 
}$$

Before stating the next definition, 
it should be reminded that $\cN_{D(L,E)}=\cN_L+\cN_E
=\cN_L^{\Om}+\cN_L^{\cX}+\cN_E$.

\begin{defn}[LRR-rewrite rule]
A \emph{Local Redirection and Replacement Rewrite rule}
(or a \emph{LRR-rewrite rule}, for short)
is a span of graph homomorphisms of the form:
  $$ \xymatrix@C=5pc{ L & D(L,E) \ar[l]_{\delta_{L,E}} \ar[r]^{\rho} &
   R }$$
where $E$ is a set of edges of $L$,
and where $\rho(\cN_L^{\cX})\subseteq\cN_R^{\cX}$
and the restriction of $\rho$ to $\cN_L^{\cX}$ is injective.
\end{defn}

\begin{exam}
  \label{singleton} Consider the function $add$ which adds an element
  to a \emph{circular} list. The span below defines a rewrite rule
  defining the function $add$ in the case where the circular list
  consists of one element (the case of lists of length greater than one is
given in Example~\ref{add2}).

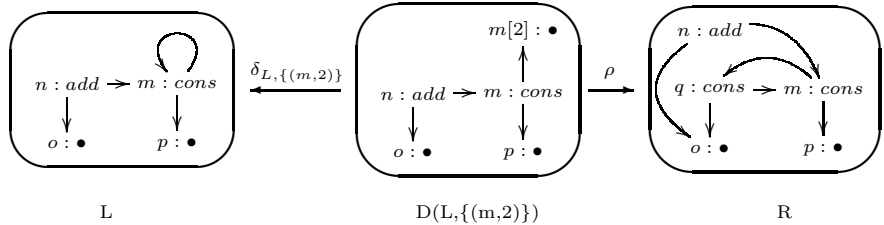
\newcommand{\LCIRCUN}{
   \xymatrix@R=1pc@C=.7pc{
    \\
    n:add \ar[d]\ar[r]& m:cons \ar[d]\ar@(ur,ul)[]\\
   o:\variable & p:\variable  
   }}

\newcommand{\DCIRCUN}{
  \xymatrix@R=1pc@C=.7pc{
    & m[2]:\variable \\
    n:add \ar[d]\ar[r]& m:cons \ar[d]\ar[u] \\
    o:\variable & p:\variable}}

\newcommand{\RCIRCUN}{
  \xymatrix@R=1pc@C=.7pc{
    n:add \ar@/^1pc/[dr]\ar@/^-1.7pc/[dd]\\
    q:cons \ar[d]\ar[r] & m: cons \ar[d] \ar@/^-1pc/[l] \\ 
    o:\variable & p:\variable
     }}

\scriptsize
\setlength{\unitlength}{1mm}
\begin{picture}(160,30)
  \put(-2,25){\Ovalbox{\LCIRCUN}} \put(10,0){L}
  \put(44,25){\Ovalbox{\DCIRCUN}} \put(52,0){D(L,\{(m,2)\})}
  \put(83,25){\Ovalbox{\RCIRCUN}} \put(100,0){R}
 
  \linethickness{.8pt}
  \put(42,17){\vector(-1,0){12}} \put(30,19){$\delta_{L,\{(m,2)\}}$}
  \put(75,17){\vector(1,0){6}}  \put(77,19){$\rho$}
\end{picture}
\normalsize
 
  In this example we show how (local) edge redirection can be achieved
  through edge disconnection. Since an element is added to the head of
  a circular list (of length 1), one has to make the curve pointer
  $(m,2)$ to point to the new added cell. For this we disconnect the
  edge $(m,2)$ in $D(L,\{(m,2)\})$ in order to be able to redirect it,
  thanks to an appropriate homomorphism $\rho$, to the new cell in
  $R$, namely $q$. Here, $\rho = [n \mapsto n ; m[2] \mapsto q; \cdots ]$

  One may also remark that graph $R$ still has a node labelled by
$add$. In this paper we do not tackle the problem of garbage
collection which has been treated in a categorical way in e.g. \cite{Banach94}.

\end{exam}

\begin{defn}[LRR-matching]
 \label{defn:match}
A \emph{LRR-matching} with respect to a LRR-rewrite rule 
$\xymatrix@C=2pc{L & D(L,E) \ar[l]_{\delta_{L,E}} \ar[r]^{\rho} & R }$ 
is a graph homomorphism $\mu:L\to U$ that is \emph{$\Om$-injective},
which means that the restriction of the map $\mu$ to $\cN_G^{\Om}$
is injective.
\end{defn}

\begin{defn}[LRR-Rewrite step]
Let $r=(\xymatrix@C=2pc{ 
L & D(L,E) \ar[l]_{\delta_{L,E}} \ar[r]^{\rho} & R })$
be a rewrite rule, and $\mu:L\to U$ a matching with respect to~$r$.
Then $U$ \emph{rewrites into $V$ using rule $r$} if 
there are graph homomorphisms $\nu:R\to V$ and 
$\rho':D(U,\mu(E))\to V$ such that 
the following square is a pushout in the category of graphs ($\Gr$):
$$ \xymatrix@C=5pc{
D(L,E) \ar[d]_{D_{\mu,E}} \ar[r]^{\rho} & R \ar[d]^{\nu} \\ 
D(U,\mu(E)) \ar[r]_{\quad\rho'} & V \\ 
}$$
\end{defn}

Thus, a rewrite step corresponds to a \emph{double pushout} 
in the category of graphs:
$$ \xymatrix@C=5pc{
L \ar[d]_{\mu} & D(L,E) \ar[d]_{D_{\mu,E}} \ar[l]_{\delta_{L,E}} \ar[r]^{\rho} 
& R \ar[d]^{\nu} \\ 
U & D(U,\mu(E)) \ar[l]^{\delta_{U,\mu(E)}\quad} \ar[r]_{\quad\rho'} & V \\
}$$

\begin{thm}[Rewrite step is feasible]
\label{thm:dirpo}
Let $r$ be a rewrite rule, and $\mu:L\to U$ a matching with respect to $r$.
Then $U$ can be rewritten using rule $r$.
More precisely, the required pushout can be built as follows
(the notations are simplified by dropping $E$ and $\mu(E)$): 
\begin{itemize}
\item the set of nodes of $V$ is $\cN_V=(\cN_R+\cN_{D(U)})/\sim$,
where $\sim$ is the equivalence relation generated by 
$D_{\mu}(n)\sim\rho(n)$ for each node $n$ of $D(L)$,
\item the maps $\nu$ and $\rho'$, on the sets of nodes, 
are the inclusions of $\cN_R$ and $\cN_{D(U)}$ in 
$\cN_R+\cN_{D(U)}$, respectively, followed by the quotient map
with respect to $\sim$,
\item $\cN_V^{\Om}$ is made of the classes 
modulo $\sim$ which contain at least one labeled node,
and a section $\pi:\cN_V^{\Om}\to\cN_R^{\Om}+\cN_{D(U)}^{\Om}$ 
of the quotient map is chosen, which means that 
the class of $\pi(n)$ is $n$, for each $n\in\cN_V^{\Om}$,
\item for each $n\in\cN_V^{\Om}$,
the label of $n$ is the label of $\pi(n)$,
\item for each $n\in\cN_V^{\Om}$,
the successors of $n$ are the classes of the successors
of $\pi(n)$.
\end{itemize}
Moreover, the resulting pushout does not depend on the choice of 
the section $\pi$.
\end{thm} 


\begin{coro}[A description of the labeled nodes]
\label{coro:dirpo}
With the notations and assumptions of Theorem~\ref{thm:dirpo},
the representatives of the equivalence classes of nodes of $\cN_R+\cN_{D(U)}$ 
can be chosen in such a way that:
  $$ \cN_V^{\Om} = (\cN_U^{\Om}-\mu(\cN_L^{\Om}))+\cN_R^{\Om} \;.$$
\end{coro}

\begin{dproof} 
Both Theorem~\ref{thm:dirpo} and Corollary~\ref{coro:dirpo}
are derived from Theorem~\ref{thm:grpo},
their proofs are given at the end of the appendix.
\end{dproof}

\begin{exam}
   \label{exam:inj}
   Here we consider the case of a non $\Om$-injective matching in order to
show that there may be no double pushout in such cases. Thus
justifying our restriction over acceptable matchings (see Definition~\ref{defn:match}).

  In this example we identify two nodes of $L$ labelled by $g$ via the
  homomorphism $\mu$, namely $n_1$ and $n_2$, to a single one, $m$. In
  the span we disconnect the two edges coming from $g$'s and redirect
  them to two different nodes labeled by different constants : $b$ and
  $c$.This is done by the homomorphism $\rho = id$. Now, as both edges
  have been merged by the matching in $U$, the second (right) pushout
  cannot exist since a single edge cannot point to both $b$ and $c$ in the
  same time. Note that this impossibility does not denote a limitation
  of our formalism. 
 
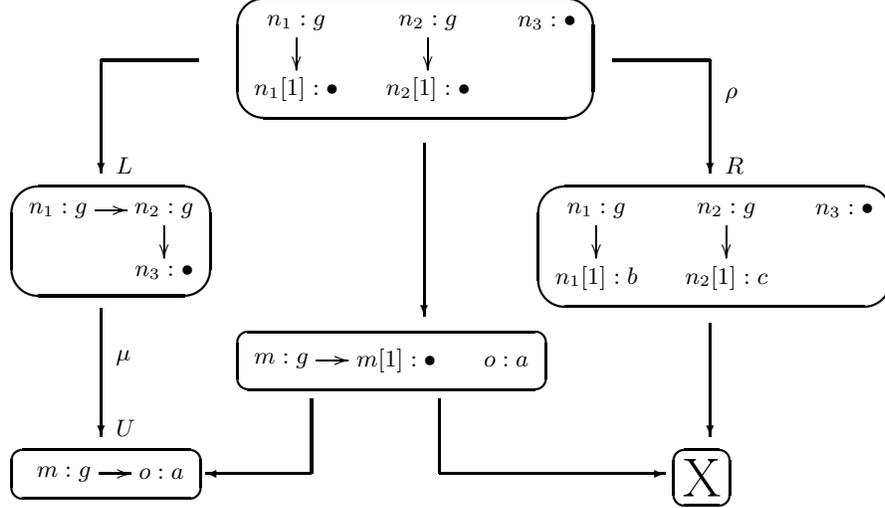
\newcommand{\LINJ}{\xymatrix@R=1pc@C=1pc
  { n_1:g \ar[r] & n_2:g \ar[d]\\
    & n_3:\variable }}

\newcommand{\DINJ}{\xymatrix@R=1pc@C=1pc
  { n_1:g \ar[d] & n_2:g \ar[d] & n_3:\variable \\
    n_1[1]:\variable & n_2[1]:\variable}}

\newcommand{\RINJ}{\xymatrix@R=1pc@C=1pc
  { n_1:g \ar[d] & n_2:g \ar[d] & n_3:\variable \\
  n_1[1]:b &n_2[1]:c}}

\newcommand{\UINJ}{ \xymatrix@R=1pc@C=1pc
  {m:g \ar[r]& o:a}}

\newcommand{\VINJ}{\xymatrix@R=1pc@C=1pc
  {m:g \ar[r] & m[1]:\variable & o:a }}

   \setlength{\unitlength}{1mm}

\small
\begin{picture}(140,66)(10,-8)

   \linethickness{.8pt}
   \put(24,35){$L$}
   \put(105,35){$R$}
   \put(24,0){$U$}
   \put(105,45){$\rho$}
   \put(24,10){$\mu$}

   \put(10,30){\Ovalbox{\LINJ}}
      \put(22,17){\vector(0,-1){17}}
   \put(40,55){\Ovalbox{\DINJ}} 
       \put(22,50){\vector(0,-1){15}} \put(22,50){\line(1,0){13}}
       \put(65,39){\vector(0,-1){23}}
       \put(103,50){\vector(0,-1){15}} \put(103,50){\line(-1,0){13}}
   \put(80,30){\Ovalbox{\RINJ}}
       \put(103,15){\vector(0,-1){15}}
   \put(10,-5){\Ovalbox{\UINJ}}
       \put(50,-5){\vector(-1,0){14}} \put(50,-5){\line(0,1){10}}
       \put(67,-5){\vector(1,0){30}} \put(67,-5){\line(0,1){10}}
   \put(40,10){\Ovalbox{\VINJ}}
   \put(98,-8){\Ovalbox{{\huge X}}}

\end{picture}
\normalsize

\end{exam}
\begin{exam}
\label{add2}
  In this example we complete the definition of the addition of an element
to a circular list started in Example~\ref{singleton} where we gave 
a span for the case of list of size 1. In Figure~1 
 we give the span for lists of size greater than 1, as well as the application
 of the rule to a list of size 3.

  Notice how the disconnection is actually used in order to redirect
the pointer $(n_6,2)$. The homomorphisms of the bottom layer show that the
disconnected edge, pointing to the unlabeled node $c_4[2]$ is
mapped to $ c_1$ to the left and to $n_8$ to the right. The mechanism
of disconnection allows the categorical manipulation of an edge. 

The $\Om$-injectivity hypothesis is also useful in this
rule since edges $(n_6,2)$ and $(n_3,2)$ must be different, thus a
list of size less than or equal to one cannot be  matched by this rule. 

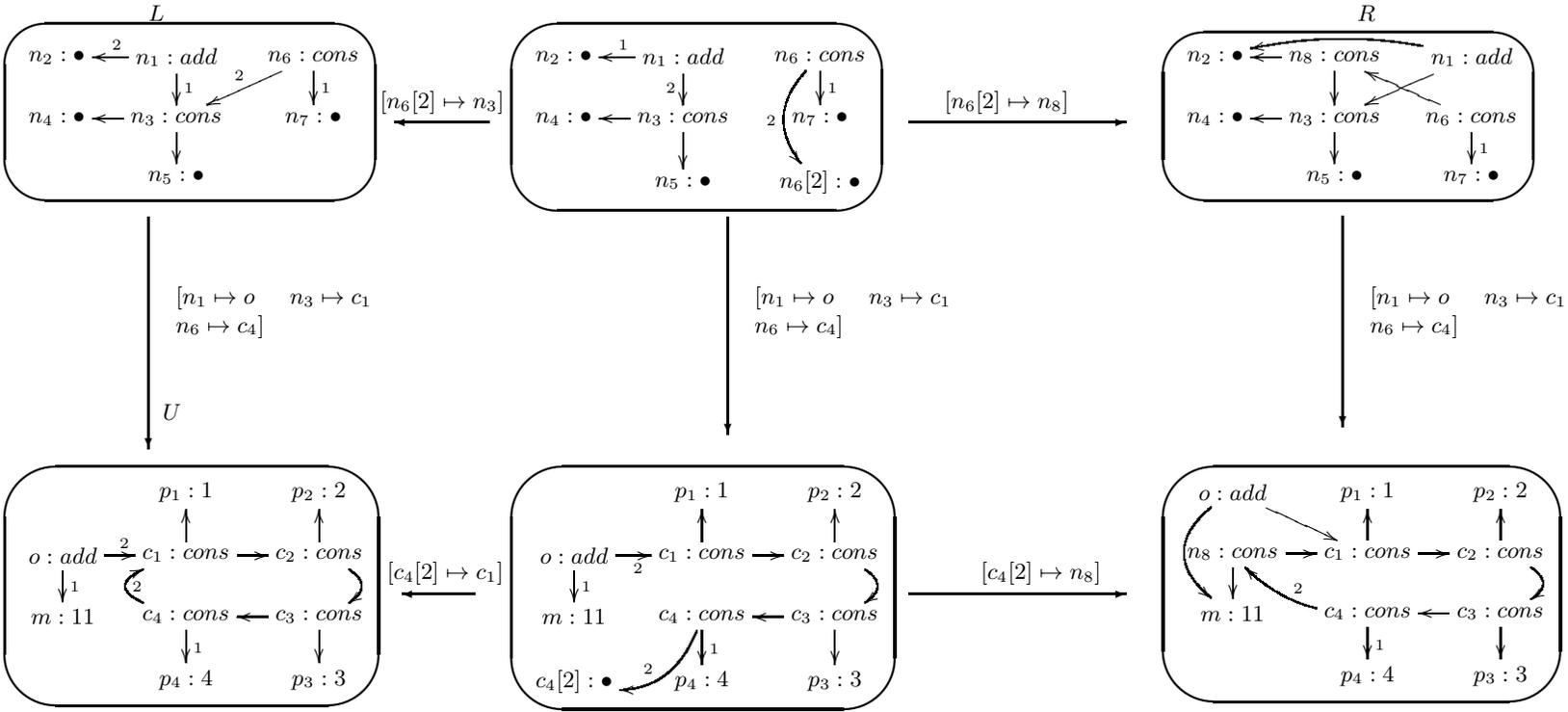
\newcommand{\LLCIRC}{
\xymatrix@R=1pc@C=1pc
  { n_2: \variable & n_1: add \ar[l]_{2} \ar[d]^{1} & n_6:cons \ar[d]^{1}\ar[dl]_{2}\\
    n_4: \variable & n_3: cons \ar[l] \ar[d] & n_7 : \variable  \\ 
    & n_5: \variable 
  }}

\newcommand{\DLCIRC}{
\xymatrix@R=1pc@C=1pc
  { n_2: \variable & n_1: add \ar[l]_{1}\ar[d]_{2} & n_6:cons \ar[d]^{1}\ar@/^-1.2pc/[dd]_{2}\\
    n_4: \variable & n_3: cons \ar[l] \ar[d] & n_7 : \variable  \\ 
    & n_5: \variable & n_6[2]: \variable
  }}

\newcommand{\RLCIRC}{
\xymatrix@R=1pc@C=1pc
  { n_2: \variable & n_8: cons \ar[l] \ar[d]& 
         n_1 : add \ar@/^-.6pc/[ll] \ar[dl] \\
    n_4: \variable & n_3: cons \ar[l] \ar[d] &n_6:cons\ar[ul]\ar[d]^{1} \\ 
   & n_5: \variable & n_7:\variable
  }}

\newcommand{\ULCIRC}{
\xymatrix@R=1pc@C=1pc
  {  &  p_1:1    & p_2 :2 \\
    o:add \ar[r]^{2} \ar[d]^{1}& c_1:cons \ar[u]\ar[r]& c_2:cons \ar[u] \ar@/^1.4pc/[d]\\
    m:11 & c_4:cons  \ar@/^2pc/[u]_{2} \ar[d]^{1}& c_3:cons \ar[l] \ar[d]\\
         &  p_4:4  & p_3:3  }}

\newcommand{\VLCIRC}{
\xymatrix@R=1pc@C=1pc
  { &  p_1:1    & p_2 :2 \\
    o:add \ar[r]_{2} \ar[d]^{1}& c_1:cons \ar[u]\ar[r]& 
      c_2:cons \ar[u] \ar@/^1.4pc/[d]\\
    m:11 & c_4:cons  \ar@/^1.2pc/[dl]_{2} \ar[d]^{1}& c_3:cons \ar[l] \ar[d]\\
      c_4[2]:\variable   &  p_4:4  & p_3:3 
  }}

\newcommand{\WLCIRC}{
\xymatrix@R=1pc@C=1pc
  {o:add \ar@/^-1.6pc/[dd] \ar[dr]&  p_1:1    & p_2 :2 \\
    n_8:cons \ar[r] \ar[d]& c_1:cons \ar[u]\ar[r]& c_2:cons 
    \ar[u] \ar@/^1.4pc/[d]\\
    m:11 & c_4:cons \ar@/^.6pc/[ul]_{2} \ar[d]^{1}& c_3:cons \ar[l] \ar[d]\\
        &  p_4:4  & p_3:3 }}

\begin{figure}
 \small

 \begin{turn}{90}
   \setlength{\unitlength}{1mm}
   \begin{picture}(200,90)(0,-10)

     \linethickness{.8pt}

     \put(22,10){$U$}
     \put(20,65){$L$}
     \put(187,65){$R$}
     \put(130,53){$[n_6[2]\mapsto n_8]$}    
     \put(52,53){$[n_6[2]\mapsto n_3]$}

     \put(0,60){\Ovalbox{\LLCIRC}} \put(67,51){\vector(-1,0){13}}
     \put(70,60){\Ovalbox{\DLCIRC}}\put(125,51){\vector(1,0){30}}
     \put(160,60){\Ovalbox{\RLCIRC}} 

     \put(20,38){\vector(0,-1){32}} 
          \put(22,24){$\begin{array}{ll} [n_1 \mapsto o & n_3 \mapsto c_1 \\
                                         n_6 \mapsto c_4]
                       \end{array}$}
     \put(100,38){\vector(0,-1){30}} 
          \put(102,24){$\begin{array}{ll}
					[n_1 \mapsto o & n_3 \mapsto c_1 \\
                                         n_6 \mapsto c_4]  
                        \end{array}$}
     \put(185,38){\vector(0,-1){29}} 
     \put(187,24){$\begin{array}{ll}
					[n_1 \mapsto o & n_3 \mapsto c_1 \\
                                        n_6 \mapsto c_4]  
                     \end{array} $}

     \put(0,0,){\Ovalbox{\ULCIRC}} 
         \put(65,-14){\vector(-1,0){10}}
         \put(53,-12){$[c_4[2] \mapsto c_1]$} 
     \put(70,0){\Ovalbox{\VLCIRC}} 
         \put(125,-14){\vector(1,0){30}} 
         \put(135,-12){$[c_4[2] \mapsto n_8]$} 
     \put(160,0){\Ovalbox{\WLCIRC}}
   \end{picture}
  \end{turn}

\normalsize
\label{dbpo}
\caption{LRR-rewrite step defining ``add'' function on circular lists of size greater than one}
\end{figure}
\end{exam}
\subsection{GR-Rewriting}

Let $U$ be graph and let $a,b \in \cN_U$.
we say that $U$ rewrites into $V$ using the global redirection
from $a$ to $b$ 
and write $U \stackrel{a\to b}{\longrightarrow} V$
iff $V$ is obtained from $U$ by redirecting all edges 
targeting node $a$ to point towards node $b$. 
This kind of rewriting is very useful when dealing
with rooted term graphs  (see, e.g. \cite{Banach94}). 
We define below one GR-rewriting step following the
double pushout approach. 

\begin{defn}[GR-rewrite rule]
A \emph{Global Redirection rewrite rule}
(or a \emph{GR-rewrite rule}, for short)
is a span of graph homomorphisms of the form:
  $$ \xymatrix@C=5pc{ P & \sw \ar[l]_{\lambda} \ar[r]^{\rho} &
   P }$$
where 
\begin{itemize}
\item $P$ is made of two unlabeled nodes $ar$ and $pr$,
\item $\sw$ (switch graph) is made of three unlabeled nodes $ar$, $pr$
and $mr$,
\item $\lambda(ar)=\lambda(mr)= ar$ and $\lambda(pr) = pr$, 
\item $\rho(ar) = ar $ and $\rho(pr)= \rho(mr) = pr$.
\end{itemize}
\end{defn}

\begin{defn}[GR-matching]
A \emph{GR-matching} with respect to a GR-rewrite rule 
$\xymatrix@C=2pc{P & \sw \ar[l]_{\lambda} \ar[r]^{\rho} &
   P }$
is a graph homomorphism $\mu:P\to U$.
\end{defn}

In order to define one GR-rewrite step,
$U \stackrel{a\to b}{\longrightarrow} V$, we need first somme technical
definitions and properties we give below.
 
\begin{defn}[Disconnected graph w.r.t. a node]
Let $G$ be a graph and  $o$ a node  of $G$.
Let $mr$ denote a node which is not in $\cN_G$.
The \emph{disconnected graph} associated to $G$ and $o$
is the following graph $\bar{D}(G,o)$:
\begin{itemize}
\item $\cN_{\bar{D}(G,o)}=\cN_G+ \{mr\}$,
\item $\cN_{\bar{D}(G,o)}^{\Om}=\cN_G^{\Om}$,
\item $\forall n\in \cN_G^{\Om}$, $\cL_{\bar{D}(G,o)}(n)=\cL_G(n)$,
\item $\forall n\in \cN_G^{\Om}$, $\forall i\in\{1,\dots,\ari(n)\}$, 
$\suc_G (n,i) = o \Rightarrow \suc_{\bar{D}(G,o)}(n,i)= mr$
\item $\forall n\in \cN_G^{\Om}$, $\forall i\in\{1,\dots,\ari(n)\}$, 
$\suc_G (n,i) \not= o \Rightarrow \suc_{\bar{D}(G,o)}(n,i)=\suc_G(n,i)$
\end{itemize}
\end{defn}

Informally, $\bar{D}(G,o)$ is obtained from the graph $G$ after
redirecting all incoming edges of node $o$ to point to the new
unlabeled node $mr$.

\begin{prop}[Inverse pushout]
Let $U$ be a graph, 
$ \xymatrix@C=2pc{ P & \sw \ar[l]_{\lambda}
\ar[r]^{\rho} & P }$ be a GR-rewrite rule,
and $\mu:P\to U$ a GR-matching.
Let $\bar{D}_{\mu}: \sw \to \bar{D}(U,\mu(ar))$ be the homomorphism defined
by $\bar{D}_{\mu}(ar) = \mu(ar)$, $\bar{D}_{\mu}(pr) = \mu(pr)$ and 
$\bar{D}_{\mu}(mr) = mr$.
Let $\delta_{\mu}:\bar{D}(U,\mu(ar)) \to U $ be the homomorphism defined
by $\delta_{\mu} (n) = n $ if $n \ne mr$ and 
$ \delta_{\mu}(mr) = \mu(ar)$.
Then the following square is a pushout in
the category of graphs ($\Gr$):
$$ \xymatrix@C=5pc{
P \ar[d]_{\mu} & \sw \ar[d]^{\bar{D}_{\mu}} \ar[l]_{\lambda} \\ 
U & \bar{D}(U,\mu(ar)) \ar[l]^{\delta_{\mu}} \\ 
}$$
\end{prop}

\begin{dproof} 
This result is a direct consequence of Theorem~\ref{thm:grsetpo}.
\end{dproof}

\begin{defn}[GR-rewrite step]
Let $U$ be a graph, 
$r =  \xymatrix@C=2pc{ P & \sw \ar[l]_{\lambda}
\ar[r]^{\rho} & P }$ be a GR-rewrite rule,
and $\mu:P\to U$ be a GR-matching. 
Let $\bar{D}_{\mu}: \sw \to \bar{D}(U,\mu(ar))$ be the homomorphism defined
by $\bar{D}_{\mu}(ar) = \mu(ar)$, $\bar{D}_{\mu}(pr) = \mu(pr)$ and 
$\bar{D}_{\mu}(mr) = mr$.
Then $U$ \emph{rewrites into $V$ using rule $r$} if 
there are graph homomorphisms $\nu: P\to V$ and 
$\rho':\bar{D}(U,\mu(ar))\to V$ such that 
the following square is a pushout in the category of graphs ($\Gr$):
$$ \xymatrix@C=5pc{
\sw \ar[d]_{\bar{D}_{\mu}} \ar[r]^{\rho} & P \ar[d]^{\nu} \\ 
\bar{D}(U,\mu(ar)) \ar[r]_{\quad\rho'} & V \\ 
}$$
\end{defn}

Thus, a GR-rewrite step, 
$U  \stackrel{\mu(ar)\to \mu(pr)}{\longrightarrow} V$, 
 corresponds to a \emph{double pushout} 
in the category of graphs:
$$ \xymatrix@C=5pc{
P \ar[d]_{\mu} & \sw \ar[d]_{\bar{D}_{\mu}} \ar[l]_{\delta_{\mu}} \ar[r]^{\rho} 
& P \ar[d]^{\nu} \\ 
U & D(U,\mu(ar)) \ar[l]^{\delta_{\mu}\quad} \ar[r]_{\quad\rho'} & V \\
}$$

The construction of graph $V$ is straightforward. It 
may be deduced from Theorem~\ref{thm:grpo} given in the appendix.

\begin{exam}
  \label{exam:globrewrite} In this example we show how global
  redirection works. In the graph $G$, given in
  Example~\ref{graphexample}, we want  redirect all edges with
  target $n$ towards $q$.  For this pupose, we define the homomorphism
  $\mu$ from $P$ to $G$ by mapping appropriately the nodes $ar$ (\emph{ante-rewriting}), and $pr$
  (\emph{post-rewriting}). I.e. in our
  case $\mu = [ar \mapsto n; pr \mapsto q]$.  Applying this on $G$, we get
  the following double push-out:

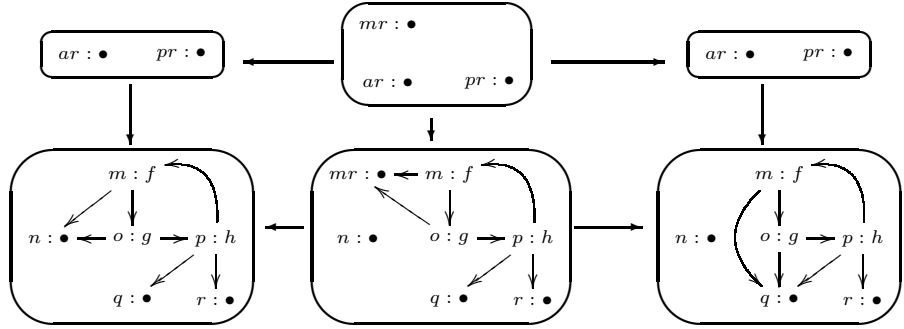
\newcommand{\GGR}
{\xymatrix@R=1pc@C=.7pc
     {&m:f\ar[dl] \ar[d]\\
      n:\variable& o:g \ar[l] \ar[r] & p:h \ar[d] \ar[dl] \ar@/^-1.4pc/[ul] \\
      & q:\variable & r:\variable}}

\newcommand{\MGR}
{\xymatrix@R=1pc@C=.7pc
     {mr:\variable & m:f\ar[l] \ar[d]\\
      n:\variable  & o:g \ar[ul] \ar[r] & p:h \ar[d] \ar[dl] \ar@/^-1.4pc/[ul] \\
      & q:\variable & r:\variable}}

\newcommand{\RGR}
{\xymatrix@R=1pc@C=.7pc
     { &m:f\ar[d] \ar@/^-1.4pc/[dd]\\
      n:\variable& o:g \ar[d] \ar[r] & p:h \ar[d] \ar[dl] \ar@/^-1.4pc/[ul] \\
      & q:\variable & r:\variable}}

\scriptsize
\setlength{\unitlength}{1mm}
\begin{picture}(160,45)(15,-20)
  \linethickness{.8pt}

  \put(10,0){\Ovalbox{\GGR}} \put(49,-7){\vector(-1,0){5}}
  \put(50,0){\Ovalbox{\MGR}} \put(85,-7){\vector(1,0){9}}
  \put(96,0){\Ovalbox{\RGR}}

  \put(26,12){\vector(0,-1){8}}
  \put(66,7.5){\vector(0,-1){3}}
  \put(110,12){\vector(0,-1){8}}

  \put(14,16){\Ovalbox{\xymatrix@R=1pc@C=1pc{ar:\variable & pr:\variable}}}
      \put(53,15){\vector(-1,0){12}}
  \put(54,20){\Ovalbox{\xymatrix@R=1pc@C=1pc{mr:\variable \\
                                  ar:\variable & pr:\variable}}}
      \put(82,15){\vector(1,0){15}}
  \put(100,16){\Ovalbox{\xymatrix@R=1pc@C=1pc{ar:\variable & pr:\variable}}}
\end{picture} 
\normalsize

Notice how node $mr$ (\emph{midrewriting}) is used. It is mapped
to $n$ on the left and to $q$ on the right. Thus in the middle graph,
 $mr$ allows to disconnect edges targeting $n$ in order to
redirect them towards $q$. 
\end{exam}
\begin{exam}

In this additional example, we give rewriting rules defining  the 
function length (written
$\sharp$) which computes the size of non-empty circular lists.
In this example  every LRR-rewriting is followed by a GR-rewriting.
That is why we precise the global rewriting that should be performed
after each LRR-rewrite step.

The first rule simply introduces an auxiliary function,
$\sharp_{b}$, which has two arguments. The first one indicates the head
of the list while the second one will move along the list in order to
measure it. We have the following span for $\sharp$:

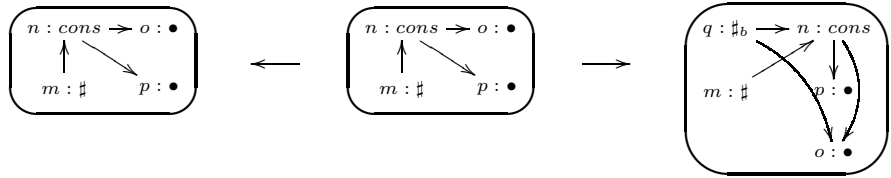
\newcommand{\LENGTHCIRC}
           {\xymatrix@R=1pc@C=.7pc{
            n:cons \ar[dr]\ar[r] & o:\variable \\
            m:\sharp \ar[u] &p:\variable}}

\scriptsize
\begin{center}
\begin{tabular}{lrclr}
   \Ovalbox{\LENGTHCIRC} & \xymatrix@R=.6pc@C=1.5pc{ \\ ~& \ar[l] \\ } &
  \Ovalbox{\LENGTHCIRC} & \xymatrix@R=.6pc@C=1.5pc{ \\ \ar[r] & ~  \\} &
  \Ovalbox{\xymatrix@R=1pc@C=1pc
           {q: \sharp_{b} \ar[r] \ar@/^.8pc/[ddr]& n:cons \ar[d]\ar@/^.8pc/[dd] \\
            m:\sharp \ar[ur] & p:\variable\\
            & o:\variable }} 
\end{tabular}
\end{center}
\normalsize
together with the pair $(m,q)$ for the global redirection.

Now we have two rules for $\sharp_b$. The first one considers
the case where the two arguments of  $\sharp_b$ are the same ; and thus
the length of the list equals one ($succ(0)$). Thus we have the following 
 span:

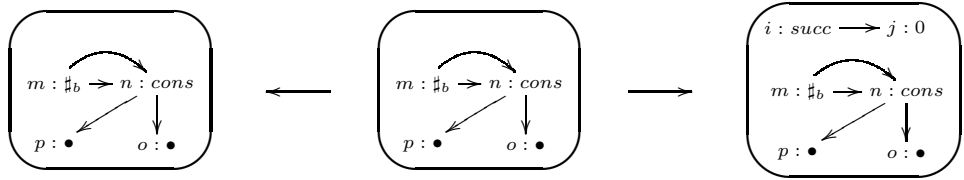
\newcommand{\LENGTHCIRCBIS}
   {\xymatrix@R=1pc@C=.7pc{ \\
    m: \sharp_b \ar@/^1pc/[r]\ar[r]& n:cons \ar[dl] \ar[d]\\
	 p:\variable & o:\variable}}

\scriptsize
\begin{center}
\begin{tabular}{lrclr}
   \Ovalbox{\LENGTHCIRCBIS}& \xymatrix@R=1.5pc@C=2pc{ \\ ~& \ar[l] \\ } &
   \Ovalbox{\LENGTHCIRCBIS}& \xymatrix@R=1.5pc@C=2pc{ \\ \ar[r] & ~  \\} &
   \Ovalbox{\xymatrix@R=1pc@C=.7pc{
    i:succ \ar[r] & j:0\\
    m: \sharp_b \ar@/^1pc/[r]\ar[r]& n:cons \ar[dl] \ar[d]\\
   p:\variable & o:\variable}}
\end{tabular}
\end{center}
\normalsize
together with the pair $(m,i)$ for the global redirection. Notice
that in this particular case we simply drop the input 
and replace it by a new graph as in classical term rewrite systems,
before performing the global redirection induced by the pair $(m,i)$.

The next  rule defines $\sharp_b$ when its arguments are different.
Once again we use the hypothesis of $\Om$-injectivity to ensure that
both $cons$ nodes cannot be identified via matching.

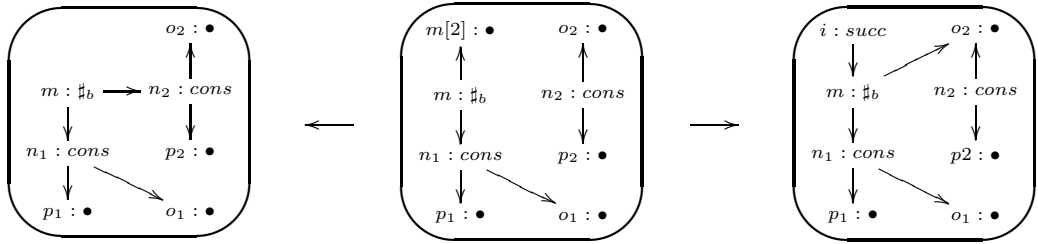
\scriptsize
\begin{center}
\begin{tabular}{lrclr}
\Ovalbox{\xymatrix@R=1pc@C=.7pc
    {& o_2 : \variable  \\
    m:\sharp_{b} \ar[r]\ar[d] & n_2:cons \ar[d]\ar[u] \\
    n_1:cons \ar[dr]\ar[d]& p_2:\variable \\
    p_1:\variable & o_1: \variable} } &
    \xymatrix@R=1pc@C=1.5pc{ \\ \\ ~& \ar[l] \\ } &
\Ovalbox{\xymatrix@R=1pc@C=.7pc
   {m[2]:\variable & o_2 : \variable\\
    m:\sharp_{b} \ar[u]\ar[d]& n_2:cons \ar[d]\ar[u]  \\
     n_1:cons \ar[dr]\ar[d] & p_2:\variable \\
     p_1:\variable  & o_1:\variable} } &
     \xymatrix@R=1pc@C=1.5pc{ \\ \\ \ar[r] & ~  \\} &
\Ovalbox{\xymatrix@R=1pc@C=.7pc
   {i:succ \ar[d] & o_2 : \variable \\
   m:\sharp_{b} \ar[ur]\ar[d] & n_2:cons \ar[d]\ar[u]\\
     n_1:cons \ar[dr]\ar[d]  & p2:\variable  \\
     p_1:\variable & o_1:\variable }}
\end{tabular}
\end{center}
\normalsize

together with the pair $(m,i)$ for the global redirection. 
We let the reader check that circular lists of size $n$ actually reduce
to $\stackrel{n}{\overbrace{succ(succ\ldots}}(0))$
by successive application of rewriting rules (LRR and GR rewrite steps).

\end{exam}

\section{Related Work}

Term graph rewriting \cite{BVG87,Plu98a,BS99} have been
mainly motivated by implementation issues of functional programming
languages. These motivations impact clearly their definition. 

In \cite{HKD91,CorradiniR93} jungles, a representation of acyclic term
graphs by means of hypergraphs, have been investigated. We share with
these proposals the use of the double-pushout approach of rewriting.
However, we are rather interested in cyclic graphs.

In \cite{BVG87,KKS94,EcJ98a} cyclic term graph rewriting is considered
using the algorithmic way. Pointer redirection  is limited to global
redirection of all edges pointing to the root of a redex by  redirecting them
to point to the root of the instance of the right-hand side.
In \cite{Banach94}, Banach, inspired by features found in
implementations of declarative languages, proposed rewrite systems
close to ours.  We share the same graphs and global redirection of
pointers. However, Banach did not discuss local redirections of
pointers. We differ also in the way to express rewriting. Rewriting
steps in \cite{Banach94} are defined by using the notion of opfibration
of a category while our approach is based on double-pushouts.

The difference between our proposal to generalize term graph rewriting
and previous works comes from the motivation.  Our aim is not the
implementation of declarative programming languages.  It is rather the
investigation of the elementary transformation rules of
data-structures as occur in classical algorithms.  In such structures
pointers play a key r\^ole that we tried to take into account by
proposing for instance redirections of specific edges within rewrite
rules.

In \cite{HabelP01}, Habel and Plump proposed a kernel language for graph
transformation. This language has been improved recently in
\cite{PlumpS04}. Basic rules in this framework are of the form $L
\leftarrow K \rightarrow R$ satisfying some conditions such as the
inclusion $K \subseteq L$. Unfortunately, our rewrite rules do not
fulfill such condition ; particularly when performing local edge
redirections. Furthermore, inverse pushouts (or pushout complements)
are not unique in our setting which is not the case in
\cite{HabelP01,PlumpS04}.

Recently, in \cite{AEH94a} the authors are also interested in
classical data-structures built by using pointers. Their work is
complementary to ours in the sense that they are rather concerned by
\emph{recognizing} data-structure shapes  by means of so 
called ¨Graph reduction specifications¨.

Last, but not least, there are yet some programming languages which
provide graph transformation features (see,
e.g. \cite{SWZ99,ERT99,GlauertKS90,Rod98}). Our purpose in this paper
is to focus on formal definition of basic data-structure
transformation steps rather than building an entire programming
language with suitable visual syntax and appropriate evaluation
strategies.

\section{Conclusion}

We defined two basic rewrite steps dedicated to data-structure
rewriting.  The rewrite relationships induced by LRR-rewrite rules as
well as GR-rewrite rules over graphs are trickier than the classical
ones over terms (trees).  There was no room in the present paper to
discuss classical properties of the rewrite relationship induced by
the above definitions such as confluence and termination or its
extension to narrowing. However, our preliminary investigation shows
that confluence is not guaranted even for nonoverlapping rewrite
systems, and thus user-definable strategies are necessary when using
all the power of data-structure rewriting.  In addition, integration
of LRR and GR rewriting in one step is also possible and can be
helpful in describing some algorithms.

On the other hand, data-structures are better represented by means of
graphics (e.g. \cite{Rod98}). Our purpose in this paper was rather the
definition of the basic rewrite steps for data-structures. We intend
to consider syntactical issue in a future work.

\bibliographystyle{abbrv}

\begin{thebibliography}{10}

\bibitem{AL}
A.~Asperti and G.~Longo.
\newblock {\em Categories, Types and Structures. An introduction to Category
  Theory for the working computer scientist.}
\newblock M.I.T. Press, 1991.
\newblock http://www.di.ens.fr/users/longo/download.html.

\bibitem{BaaderN98}
F.~Baader and T.~Nipkow.
\newblock {\em Term rewriting and all that}.
\newblock Cambridge University Press, 1998.

\bibitem{AEH94a}
A.~Bakewell, D.~Plump, and C.~Runciman.
\newblock Checking the shape safety of pointer manipulations.
\newblock In {\em International Seminar on Relational Methods in Computer
  Science (RelMiCS 7), Revised Selected Papers, Lecture Notes in Computer
  Science 3051,Springer-Verlag}, pages 48--61, 2004.

\bibitem{Banach94}
R.~Banach.
\newblock Term graph rewriting and garbage collection using opfibrations.
\newblock {\em Theoretical Computer Science}, 131:29--94, 1994.

\bibitem{BVG87}
H.~Barendregt, M.~van Eekelen, J.~Glauert, R.~Kenneway, M.~J. Plasmeijer, and
  M.~Sleep.
\newblock Term graph rewriting.
\newblock In {\em PARLE'87}, pages 141--158. LNCS 259, 1987.

\bibitem{BS99}
E.~Barendsen and S.~Smetsers.
\newblock Graph rewriting aspects of functional programming.
\newblock In H.~Ehrig, G.~Engels, H.~J. Kreowski, and G.~Rozenberg, editors,
  {\em Handbook of Graph Grammars and Computing by Graph Transformation},
  volume~2, pages 63--102. World Scientific, 1999.

\bibitem{BookF93}
R.~V. Book and F.~Otto.
\newblock {\em String-rewriting systems}.
\newblock Springer-Verlag, 1993.

\bibitem{CorradiniMREHL97}
A.~Corradini, U.~Montanari, F.~Rossi, H.~Ehrig, R.~Heckel, and M.~L{\"o}we.
\newblock Algebraic approaches to graph transformation - part {I}: Basic
  concepts and double pushout approach.
\newblock In {\em Handbook of Graph Grammars}, pages 163--246, 1997.

\bibitem{CorradiniR93}
A.~Corradini and F.~Rossi.
\newblock Hyperedge replacement jungle rewriting for term-rewriting systems and
  programming.
\newblock {\em Theor. Comput. Sci.}, 109(1{\&}2):7--48, 1993.

\bibitem{EcJ98a}
R.~Echahed and J.~C. Janodet.
\newblock Admissible graph rewriting and narrowing.
\newblock In {\em Proc.~of Joint International Conference and Symposium on
  Logic Programming (JICSLP'98)}, pages 325--340. MIT Press, June 1998.

\bibitem{handbook2}
H.~Ehrig, G.~Engels, H.-J. Kreowski, and G.~Rozenberg, editors.
\newblock {\em Handbook of Graph Grammars and Computing by Graph
  Transformations, Volume 2: Applications, Languages and Tools}. World
  Scientific, 1999.

\bibitem{handbook3}
H.~Ehrig, H.-J. Kreowski, U.~Montanari, and G.~Rozenberg, editors.
\newblock {\em Handbook of Graph Grammars and Computing by Graph
  Transformations, Volume 3: Concurrency, Parallelism and Distribution}. World
  Scientific, 1999.

\bibitem{ERT99}
C.~Ermel, M.~Rudolf, and G.~Taentzer.
\newblock The {AGG} approach: language and environment.
\newblock In {\em Handbook of graph grammars and computing by graph
  transformation: vol. 2: applications, languages, and tools}, pages 551--603.
  World Scientific Publishing Co., Inc., 1999.

\bibitem{GlauertKS90}
J.~R.~W. Glauert, R.~Kennaway, and M.~R. Sleep.
\newblock {D}actl: An experimental graph rewriting language.
\newblock In {\em Graph-Grammars and Their Application to Computer Science,
  LNCS 532}, pages 378--395, 1990.

\bibitem{HKD91}
A.~Habel, H.~J. Kreowski, and D.~Plump.
\newblock Jungle evaluation.
\newblock {\em Fundamenta Informaticae}, 15(1):37--60, 1991.

\bibitem{HMD01}
A.~Habel, J.~Muller, and D.~Plump.
\newblock Double-pushout graph transformation revisited.
\newblock {\em Mathematical Structures in Computer Science}, 11, 2001.

\bibitem{HabelP01}
A.~Habel and D.~Plump.
\newblock Computational completeness of programming languages based on graph
  transformation.
\newblock In {\em FoSSaCS LNCS 2030}, pages 230--245, 2001.

\bibitem{KKS94}
J.~R. Kennaway, J.~K. Klop, M.~R. Sleep, and F.~J.~D. Vries.
\newblock On the adequacy of graph rewriting for simulating term rewriting.
\newblock {\em ACM Transactions on Programming Languages and Systems},
  16(3):493--523, 1994.

\bibitem{Plu98a}
D.~Plump.
\newblock Term graph rewriting.
\newblock In H.~Ehrig, G.~Engels, H.~J. Kreowski, and G.~Rozenberg, editors,
  {\em Handbook of Graph Grammars and Computing by Graph Transformation},
  volume~2, pages 3--61. World Scientific, 1999.

\bibitem{PlumpS04}
D.~Plump and S.~Steinert.
\newblock Towards graph programs for graph algorithms.
\newblock In {\em ICGT, LNCS 3256}, pages 128--143, 2004.

\bibitem{Rod98}
P.~Rodgers.
\newblock {A Graph Rewriting Programming Language for Graph Drawing}.
\newblock In {\em Proceedings of the 14th IEEE Symposium on Visual Languages}.
  IEEE, IEEE Computer Society Press, September 1998.

\bibitem{handbook1}
G.~Rozenberg, editor.
\newblock {\em Handbook of Graph Grammars and Computing by Graph
  Transformations, Volume 1: Foundations}. World Scientific, 1997.

\bibitem{SWZ99}
A.~Sch\"urr, A.~J. Winter, and A.~Z\"undorf.
\newblock The {PROGRES} approach: language and environment.
\newblock In {\em Handbook of graph grammars and computing by graph
  transformation: vol. 2: applications, languages, and tools}, pages 487--550.
  World Scientific Publishing Co., Inc., 1999.

\end{thebibliography}

\appendix

\section{Pushouts of graphs}

Let $\Gr$ denote the category of graphs 
and $\Set$ the category of sets. 
The \emph{node functor} $\cN:\Gr\to\Set$ 
maps each graph $G$ to its set of nodes $\cN_G$,
and each graph homomorphism $\varphi:G\to H$ to its 
underlying map on nodes $\varphi:\cN_G\to\cN_H$.
As in the rest of the paper,
this map is simply denoted $\varphi$,
and this is not ambiguous: indeed, if two graph homomorphisms
$\varphi,\psi:G\to H$ are 
such that their underlying maps are equal $\varphi=\psi:\cN_G\to\cN_H$,
then it follows directly from the definition of 
graph homomorphisms that $\varphi=\psi:G\to H$.
In categorical terms \cite{AL}, this is expressed by
the following result. 

\begin{prop}[Faithfulness]
\label{prop:faithful}
The functor $\cN:\Gr\to\Set$ is faithful.
\end{prop}

It is worth noting that this property does not hold for 
the ``usual'' directed multigraphs, where the set of successors 
of a node is unordered. 

It is well-known that the category $\Set$ has pushouts.
On the contrary, the category $\Gr$ does not have pushouts.
For instance, let us consider a span of graphs:
$$ \xymatrix@=1pc{
& G_0 \ar[dl]_{\varphi_1} \ar[dr]^{\varphi_2} & \\ 
G_1 & & G_2  \\ 
}$$ 
where $G_0$, $G_1$ and $G_2$ are made of only one node:
$n_0$ in $G_0$ is unlabeled, 
$n_1\lb a_1$ in $G_1$ and $n_2\lb a_2$ in $G_2$,
where $a_1$ and $a_2$ are distinct constants.
This span has no pushout, 
because there cannot be any commutative square of graphs
based on it.

Theorem~\ref{thm:grsetpo} below
states a sufficient condition for a commutative square of graphs 
to be a pushout,
and Theorem~\ref{thm:grpo} 
states a sufficient condition for a span 
of graphs to have a pushout, together with a construction 
of this pushout.

In the following, when $G_i$ occurs as an index, 
it is replaced by $i$. 

\begin{thm}[Pushout of graphs from pushout of sets]
\label{thm:grsetpo}
If a square $\Gamma$ of the following form in the category of graphs:
$$ 
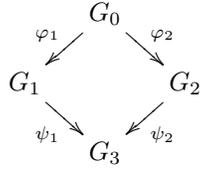
\xymatrix@=1pc{
& G_0 \ar[dl]_{\varphi_1} \ar[dr]^{\varphi_2} & \\ 
G_1 \ar[dr]_{\psi_1} & & G_2 \ar[dl]^{\psi_2} \\ 
& G_3 & \\
}$$
is such that:
\begin{enumerate}
\item $\Gamma$ is a commutative square in $\Gr$,
\item $\cN(\Gamma)$ is a pushout in $\Set$,
\item and each $n\in\cN_3^{\Omega}$ is in 
$\psi_i(\cN_i^{\Omega})$ for $i=1$ or $i=2$,
\end{enumerate}
then $\Gamma$ is a pushout in $\Gr$.
\end{thm}

Point $(2)$ implies that each $n\in\cN_3$ is the image of
at least a node in $G_1$ or in $G_2$, and point $(3)$ 
adds that, if $n$ is labeled, then it is the image of 
at least a labeled node in $G_1$ or in $G_2$.

\begin{dproof}
Let us consider a commutative square $\Gamma'$ in $\Gr$ of the form:
$$ 
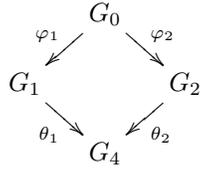
\xymatrix@=1pc{
& G_0 \ar[dl]_{\varphi_1} \ar[dr]^{\varphi_2} & \\ 
G_1 \ar[dr]_{\theta_1} & & G_2 \ar[dl]^{\theta_2} \\ 
& G_4 & \\
}$$
Then $\cN(\Gamma')$ is a commutative square in $\Set$,
and since $\cN(\Gamma)$ is a pushout in $\Set$,
there is a unique map $\theta:\cN_3\to\cN_4$
such that $\theta\circ\psi_i=\theta_i$, for $i=1,2$.
$$ 
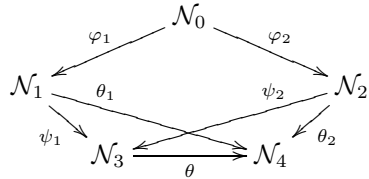
\xymatrix@=1pc{
&& \cN_0 \ar[dll]_{\varphi_1} \ar[drr]^{\varphi_2} && \\ 
\cN_1 \ar[dr]_{\psi_1} \ar[drrr]^(.3){\theta_1} &&&& 
\cN_2 \ar[dlll]_(.3){\psi_2} \ar[dl]^{\theta_2} \\ 
& \cN_3 \ar[rr]_{\theta} && \cN_4 & \\
}$$
Let us now prove that $\theta$ actually is a graph homomorphism.
According to Definition~\ref{defn:grhom}, we have to prove that,
for each labeled node $n$ of $G_3$, its image $n'=\theta(n)$
is a labeled node of $G_4$, 
and that $\cL_4(n') = \cL_3(n)$ and $\cS_4(n') = \theta^*(\cS_3(n))$.

So, let $n\in\cN_3^{\Omega}$, and let $n'=\theta(n)\in\cN_4$.
>From our third assumption, without loss of generality, 
$n=\psi_1(n_1)$ for some $n_1\in\cN_1^{\Omega}$.
It follows that
$\theta_1(n_1)=\theta(\psi_1(n_1))=\theta(n)=n'$:
  $$ n=\psi_1(n_1) \;\mbox{ and }\; n'=\theta_1(n_1) \;.$$

Since $n_1$ is labeled and $\theta_1$ is a graph homomorphism,
the node $n'$ is labeled.

Since $\psi_1$ and $\theta_1$ are graph homomorphisms,   
$\cL_3(n)=\cL_1(n_1)$ and $\cL_4(n')=\cL_1(n_1)$,
thus $\cL_3(n)=\cL_4(n')$, as required for labels.

Since $\psi_1$ and $\theta_1$ are graph homomorphisms,   
$\cS_3(n)=\psi_1^*(\cS_1(n_1))$ and $\cS_4(n')={\theta_1}^*(\cS_1(n_1))$.
So, $\theta^*(\cS_3(n))=\theta^*(\psi_1^*(\cS_1(n_1)))
={\theta_1}^*(\cS_1(n_1)=\cS_4(n')$, as required for successors.

This proves that $\theta:G_3\to G_4$ is a graph homomorphism.
Then, from the faithfulness of the functor $\cN$ 
(Proposition~\ref{prop:faithful}), 
for $i\in\{1,2\}$, the equality of the underlying maps 
$\theta\circ\psi_i=\theta_i:\cN_i\to\cN_4$
is an equality of graph homomorphisms:
$\theta\circ\psi_i=\theta_i:G_i\to G_4$.
 
Now, let $\theta':G_3\to G_4$ be a graph homomorphism such that 
$\theta'\circ\psi_i=\theta_i$ for $i\in\{1,2\}$.
Since $\cN(\Gamma)$ is a pushout in $\Set$, 
the underlying maps are equal: $\theta=\theta':\cN_3\to\cN_4$.
Then, it follows from the faithfulness of the functor $\cN$
that the graph homomorphisms are equal: $\theta=\theta':G_3\to G_4$.
\end{dproof}

For each span of graphs $\Sigma$:
$$ \xymatrix@=1pc{
& G_0 \ar[dl]_{\varphi_1} \ar[dr]^{\varphi_2} & \\ 
G_1 & & G_2 \\ 
}$$
let $\sim$ denote the equivalence relation on the disjoint union 
$\cN_1+ \cN_2$ generated by: 
$$ \varphi_1(n_0)\sim\varphi_2(n_0) \;\mbox{ for all }\; n_0\in \cN_0
\;,$$
let $N_3$ be the quotient set $N_3=(\cN_1+\cN_2)/\sim$, 
and $\psi:\cN_1+\cN_2\to N_3$ the quotient map.
Two nodes $n,n'$ in $\cN_1+\cN_2$ are called \emph{equivalent}
if $n\sim n'$.
For $i\in\{1,2\}$, let $\psi_i:\cN_i\to N_3$ be 
made of the inclusion of 
$\cN_i$ in $\cN_1+\cN_2$ followed by $\psi$.
Then, it is well-known that the square of sets:
$$ \xymatrix@=1pc{
& \cN_0 \ar[dl]_{\varphi_1} \ar[dr]^{\varphi_2} & \\ 
\cN_1 \ar[dr]_{\psi_1} & & \cN_2 \ar[dl]^{\psi_2} \\ 
& N_3 & \\
}$$
is a pushout, which can be called \emph{canonical}.

\begin{defn}[Strongly labeled span of graphs]
\label{defn:grpo}
A span of graphs:
$$ 
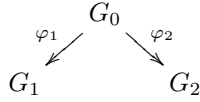
\xymatrix@=1pc{
& G_0 \ar[dl]_{\varphi_1} \ar[dr]^{\varphi_2} & \\ 
G_1 & & G_2 \\ 
}$$
is \emph{strongly labeled} if for each $n_3\in(\cN_1+\cN_2)/\sim$: 
\begin{itemize}
\item all the labeled nodes in the class $n_3$ have the 
same label, 
\item and all the labeled nodes in the class $n_3$ have 
equivalent successors.
\end{itemize}
\end{defn}

\begin{thm}[Pushout of a strongly labeled span of graphs]
\label{thm:grpo}
A strongly labeled span of graphs has a pushout:
$$ 
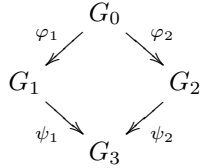
\xymatrix@=1pc{
& G_0 \ar[dl]_{\varphi_1} \ar[dr]^{\varphi_2} & \\ 
G_1 \ar[dr]_{\psi_1} & & G_2 \ar[dl]^{\psi_2} \\ 
& G_3 & \\ 
}$$
which can be built as follows:
\begin{itemize}
\item the underlying  square of sets is the canonical 
pushout square, so that $\cN_3=(\cN_1+\cN_2)/\sim$,
\item $\cN_3^{\Om}$ is made of the classes of $\cN_1+\cN_2$
(modulo $\sim$) which contain at least one labeled node,
\item for each $n_3\in\cN_3^{\Om}$,
the label of $n_3$ is the label 
of any labeled node in the class $n_3$,
\item for each $n_3\in\cN_3^{\Om}$,
the successors of $n_3$ are the classes of the successors
of any labeled node in the class $n_3$.
\end{itemize}
\end{thm} 

\begin{dproof} 
It follows easily from Theorem~\ref{thm:grsetpo}
that this square is a pushout of graphs.
\end{dproof}

\begin{proofof}{Theorem~\ref{thm:dirpo}}
(the notations are simplified by dropping $E$ and $\mu(E)$).
\\ Let us prove that the following span of graphs is strongly labeled:
$$ 
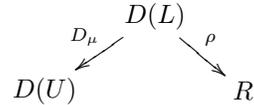
\xymatrix@=1pc{
& D(L) \ar[dl]_{D_{\mu}} \ar[dr]^{\rho} & \\ 
D(U) & & R \\ 
}$$
Then, Theorem~\ref{thm:dirpo} derives easily 
from Theorem~\ref{thm:grpo}. 

Let $n,n'\in\cN_R^{\Om}+\cN_{D(U)}^{\Om}$ be 
distinct equivalent nodes.
We have to prove that $n$ and $n'$ have the same label 
and that their successors are pairwise equivalent. 

>From the definition of the equivalence relation $\sim$, 
there is a chain of relations:
$$ 
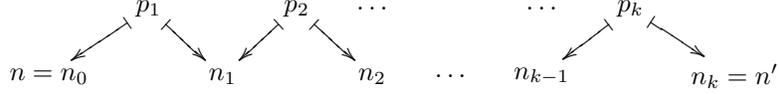
\xymatrix@C=1pc@R=1pc { & p_1 \ar@{|->}[dl]\ar@{|->}[dr] & 
& p_2 \ar@{|->}[dl]\ar@{|->}[dr] &  \dots &  & \dots &
p_k \ar@{|->}[dl]\ar@{|->}[dr] &  \\ 
n=n_0 &  & n_1 
&  & n_2 & \dots & n_{k-1} & 
 & n_k=n' \\ 
}$$
for some $k\geq1$, where each $p_i$ is in $\cN_{D(L)}$,
each $n_i$ in $\cN_{D(U)}+\cN_R$, and 
the mappings are either $D_{\mu}$ or $\rho$.
Let us assume that this chain has minimal length,
among similar chains from $n$ to $n'$.
Then:
\begin{itemize}
\item if $p_i=p_j$ for some $i<j$, the part 
of the chain between $p_i$ and $p_j$ can be dropped, 
giving rise to a shorter chain from $n$ to $n'$:
hence all the $p_i$'s are distinct;
\item if $n_{i-1}$ and $n_i$ are both in $\cN_R$, 
then $n_{i-1}=\rho(p_i)=n_i$,
and the part 
of the chain between $n_{i-1}$ and $n_i$ can be dropped, 
giving rise to a shorter chain from $n$ to $n'$:
hence $n_{i-1}$ and $n_i$ cannot be both in $\cN_R$;
\item similarly, $n_{i-1}$ and $n_i$ cannot be both in 
$\cN_{D(U)}$.
\end{itemize}

If all the nodes in this chain are labeled, then,
since $D_{\mu}$ and $\rho$ are graph homomorphisms,  
all nodes in the chain have the same label 
and have pairwise equivalent successors,
so that the result follows. 

We now prove that all the nodes in the chain are labeled,
by contradiction.
Let us assume that at least one node in the chain
is unlabeled.
Since $\rho$ and $D_{\mu}$ are graph homomorphisms,  
the first unlabeled node (starting from $n$) is some $p_i$.
Let us focus on: 
$$ 
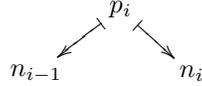
\xymatrix@C=1pc@R=1pc {
 & p_i \ar@{|->}[dl]\ar@{|->}[dr] & \\
n_{i-1} & &n_i \\
} $$
where $n_{i-1}$ is labeled and $p_i$ is unlabeled. 

It should be reminded that:
\begin{itemize}
\item $\cN_{D(L)}=\cN_L+\cN_E$ and $\cN_{D(U)}=\cN_U+\mu(\cN_E)$,
with $D_{\mu}(\cN_L)\subseteq\cN_U$ 
and $D_{\mu}$ injective on $\cN_E$
(the last point comes from the fact that  
$\mu$ is $\Om$-injective);
\item $\rho(\cN_L^{\cX})\subseteq\cN_R^{\cX}$
and the restriction of $\rho$ to $\cN_L^{\cX}$ is injective,
since $\xymatrix@C=1pc{L&D(L)\ar[l]_{\delta_{L}}\ar[r]^{\rho}&R}$
is a rewrite rule.
\end{itemize}

\textsl{Case 1: $n_{i-1}$ is a node of $R$.}
Then $n_{i-1}\in\cN_R^{\Om}$.
Since $\rho(\cN_L^{\cX})\subseteq\cN_R^{\cX}$
and $p_i$ is unlabeled, it follows that $p_i\in\cN_E$.
Then, since $D_{\mu}$ maps $\cN_E$ to $\mu(\cN_E)$, 
$n_i\in\mu(\cN_E)$. Then $k>i$, since the last node in 
the chain is labeled. 
Since $D_{\mu}$ is injective on $\cN_E$,
and maps $\cN_L$ to $\cN_U$,
it follows that $p_{i+1}=n_i$.
So, $p_i=p_{i+1}$, which is impossible 
since the chain is minimal.
$$ 
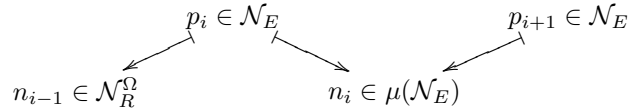
\xymatrix@C=1pc@R=1pc {
 & p_i\in\cN_E \ar@{|->}[dl]\ar@{|->}[dr] & & 
   p_{i+1}\in\cN_E \ar@{|->}[dl] \\
 n_{i-1}\in\cN_R^{\Om} & & n_i\in\mu(\cN_E) & \\
} $$

\textsl{Case 2: $n_{i-1}$ is a node of $D(U)$.}
Then $n_{i-1}\in\cN_U^{\Om}$.
Since $D_{\mu}$ maps $\cN_E$ to $\mu(\cN_E)$ 
and $D_{\mu}(\cN_L)$ on $\cN_U$,
it follows that $p_i\in\cN_L^{\cX}$.
Since $\rho$ maps $\cN_L^{\cX}$ to $\cN_R^{\cX}$,
it follows that $n_i\in\cN_R^{\cX}$.
Then $k>i$, since the last node in 
the chain is labeled.
Then $p_{i+1}\in\cN_L^{\cX}+\cN_E$.
If $p_{i+1}\in\cN_E$, a contradiction follows as in case~1.
Hence, $p_{i+1}\in\cN_L^{\cX}$.
Since the restriction of $\rho$ to $\cN_L^{\cX}$ is injective,
$p_{i+1}=p_i$, which is also impossible 
since the chain is minimal.
$$ 
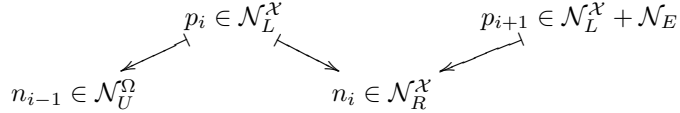
\xymatrix@C=1pc@R=1pc {
 & p_i\in\cN_L^{\cX} \ar@{|->}[dl]\ar@{|->}[dr] & & 
   p_{i+1}\in\cN_L^{\cX}+\cN_E \ar@{|->}[dl] \\
n_{i-1}\in\cN_U^{\Om} & &n_i\in\cN_R^{\cX} & \\
} $$

Finally, it has been proved that 
all the nodes in this chain are labeled, 
which concludes the proof.
\end{proofof}

\begin{proofof}{Corollary~\ref{coro:dirpo}}
We use the proof of theorem~\ref{thm:dirpo},
as well as the notations in this proof.
Let $n\in\cN_V^{\Om}$, we have to choose 
a representative $r(n)$ of $n$. 
It should be reminded that $\cN_{D(U)}^{\Om}=\cN_U^{\Om}$.

\textbf{(R.)}
If there is a node $n_R\in\cN_R^{\Om}$
such that $n=\nu(n_R)$, let us prove that it is unique.
Let $n'_R\in\cN_R^{\Om}$ be another 
node such that $n=\nu(n'_R)$, i.e., such that $n_R\sim n'_R$.
Let us consider a chain with minimal length $k\geq1$ from 
$n_R(=n_0)$ to $n'_R(=n_k)$;
we know that all the nodes in this chain are labeled.
Since $n_0$ and $n_1$ cannot be both in $\cN_R$,
it follows that $n_1\in\cN_U^{\Om}$, 
so that $p_0,p_1\in\cN_L^{\Om}$ and $n_1=\mu(p_0)=\mu(p_1)$. 
The $\Om$-injectivity of $\mu$ implies that $p_0=p_1$,
but this is impossible.
So, we have proved that $\nu^{\Om}:\cN_R^{\Om}\to\cN_V^{\Om}$
is injective, and we define $r(n)=n_R$.

\textbf{(U.)}
If there is no node $n_R\in\cN_R^{\Om}$ such that $n=\nu(n_R)$, 
then there is a node $n_U\in\cN_U^{\Om}$ such that $n=\rho'(n_U)$. 
Let us prove that it is unique. 
Let $n'_U\in\cN_U^{\Om}$ be another 
node such that $n=\rho'(n'_U)$, i.e., such that $n_U\sim n'_U$.
Let us consider a chain with minimal length $k\geq1$ from 
$n_U(=n_0)$ to $n'_U(=n_k)$;
we know that all the nodes in this chain are labeled.
Since $n_0$ and $n_1$ cannot be both in $\cN_U$,
it follows that $n_1\in\cN_R^{\Om}$, 
which contradicts our assumption: 
there is no node $n_R\in\cN_R^{\Om}$ such that $n=\nu(n_R)$. 
Let $\widetilde{\cN_U^{\Om}}$ denote the 
subset of $\cN_U^{\Om}$ made of the 
nodes which are not equivalent to any node in $\cN_R^{\Om}$.
So, we have proved that the restriction of
${\rho'}^{\Om}:\cN_{D(U)}^{\Om}\to\cN_V^{\Om}$ 
to $\widetilde{\cN_U^{\Om}}$ is injective,
and we define $r(n)=n_U$.

\textbf{(L.)}
We still have to prove that $\widetilde{\cN_U^{\Om}}=
\cN_U^{\Om}-\mu(\cN_L^{\Om})$, i.e., 
that a node $n_U\in\cN_U^{\Om}$ is equivalent to 
a node $n_R\in\cN_R^{\Om}$
if and only if there is node $n_L\in\cN_L^{\Om}$ such that $n_U=\mu(n_L)$.
\\ Clearly, if $n_L\in\cN_L^{\Om}$ and $n_U=\mu(n_L)$,
let $n_R=\rho(n_L)$, then $n_R\in\cN_R^{\Om}$ and $n_U\sim n_R$.
\\ Now, let $n_U\sim n_R$ for some $n_U\in\cN_U^{\Om}$ 
and $n_R\in\cN_R^{\Om}$. 
Let us consider a chain with minimal length $k\geq1$ from 
$n_R(=n_0)$ to $n_U(=n_k)$;
we know that all the nodes in this chain are labeled.
If $k>1$, then the $\Om$-injectivity of $\mu$ leads to a
contradiction, as in part \textbf{(R)} of the proof. 
Hence $k=1$, which means that $p_1\in \cN_L^{\Om}$ is such that
$n_R=\rho(p_1)$ and $n_U=\mu(p_1)$,
so that there is node $n_L=p_1\in\cN_L^{\Om}$ such that
$n_U=\mu(n_L)$.

This concludes the proof that:
  $$ \cN_V^{\Om} = (\cN_U^{\Om}-\mu(\cN_L^{\Om}))+\cN_R^{\Om} \;.$$
\end{proofof}

\end{document}